\documentclass{svjour3}
\pdfoutput=1

\usepackage[noblocks]{authblk}

\usepackage{graphicx}
\usepackage{array}
\usepackage{pbox}
\usepackage{wrapfig}
\usepackage{adjustbox}

\usepackage{amsmath,amssymb,amsfonts}
\usepackage{mathtools}

\usepackage{url}
\usepackage{verbatim}
\usepackage{multirow}
\usepackage{rotating}
\usepackage{todonotes}

\usepackage[linesnumbered,ruled,vlined]{algorithm2e}
\usepackage{algpseudocode}

\usepackage{ifpdf}
\usepackage{lmodern}
\usepackage{lipsum}
\usepackage[T1]{fontenc}
\usepackage{textcomp}

\usepackage{enumitem}
\usepackage{mdwlist}

\usepackage{listings}

\usepackage{comment}
\usepackage{cite}
\usepackage{ifthen}
\usepackage{color}
\usepackage{colortbl,xcolor}

\usepackage{hhline}

\usepackage{hyperref}
\usepackage{float}
\usepackage[font=small,skip=2pt]{caption}
\usepackage{subcaption}

\definecolor{LightCyan}{rgb}{0.88,1,1}
\definecolor{Gray}{gray}{0.9}
\definecolor{lightgray}{rgb}{0.83, 0.83, 0.83}
\definecolor{darkgray}{rgb}{0.66, 0.66, 0.66}
\specialcomment{notes}{\begingroup \color{blue}} { \endgroup}

\colorlet{punct}{red!60!black}
\definecolor{background}{HTML}{EEEEEE}
\definecolor{delim}{RGB}{20,105,176}
\colorlet{numb}{magenta!60!black}

\lstdefinelanguage{json}{
    basicstyle=\small\ttfamily,
    numbers=left,
    numbers=none,
    stepnumber=1,
    numbersep=8pt,
    showstringspaces=false,
    breaklines=true,
    frame=lines,
    backgroundcolor=\color{background},
    literate=
     *{0}{{{\color{numb}0}}}{1}
      {1}{{{\color{numb}1}}}{1}
      {2}{{{\color{numb}2}}}{1}
      {3}{{{\color{numb}3}}}{1}
      {4}{{{\color{numb}4}}}{1}
      {5}{{{\color{numb}5}}}{1}
      {6}{{{\color{numb}6}}}{1}
      {7}{{{\color{numb}7}}}{1}
      {8}{{{\color{numb}8}}}{1}
      {9}{{{\color{numb}9}}}{1}
      {:}{{{\color{punct}{:}}}}{1}
      {,}{{{\color{punct}{,}}}}{1}
      {\{}{{{\color{delim}{\{}}}}{1}
      {\}}{{{\color{delim}{\}}}}}{1}
      {[}{{{\color{delim}{[}}}}{1}
      {]}{{{\color{delim}{]}}}}{1},
}

\newcolumntype{!}{>{\global\let\currentrowstyle\relax}}
\newcolumntype{^}{>{\currentrowstyle}}

\DeclareGraphicsExtensions{.eps,.pdf,.png}
\graphicspath{{figure/eps/},{figure/pdf/},{figure/png/}}

\newcommand{\superscript}[1]{\ensuremath{^{\textrm{#1}}}}

\newcommand{\si}{\begin{enumerate}[leftmargin=*, itemindent=0cm, align=left]\itemsep0em}

\newcommand{\ei}{\end{enumerate}}
\begin{document}




\title{Mining Half a Billion Topical Experts Across Multiple Social Networks}
%

\def\kloutinc{\superscript{*}}


\author{Nemanja Spasojevic \and Prantik Bhattacharyya \and Adithya Rao}


\institute{Lithium Technologies | Klout \at
              San Francisco, CA \\
              \email{\{nemanja, prantik, adithya\}@klout.com}           
}


\maketitle

\begin{abstract}
Mining topical experts on social media is a problem that has gained significant attention due to its wide-ranging applications.
Here we present the first study that combines data from four major social networks -- Twitter, Facebook, Google+ and LinkedIn, along with the Wikipedia graph and internet webpage text and metadata, to rank topical experts across the global population of users.
We perform an in-depth analysis of 37 features derived from various data sources such as message text, user lists, webpages, social graphs and wikipedia. 
This large-scale study includes more than 12 billion messages over a 90-day sliding window and 58 billion social graph edges.
Comparison reveals that features derived from Twitter Lists, Wikipedia, internet webpages and Twitter Followers are especially good indicators of expertise.
We train an expertise ranking model using these features on a large ground truth dataset containing almost 90,000 labels.
This model is applied within a production system that ranks over 650 million experts in more than 9,000 topical domains on a daily basis.
We provide results and examples on the effectiveness of our expert ranking system, along with empirical validation.
Finally, we make the topical expertise data available through open REST APIs for wider use.

\end{abstract}




\section{Introduction}
\label{section:introduction}

Millions of people use online social networks as platforms for communication, creating billions of interactions daily.
The data from these social networks provide unique opportunities to understand the nature of social interactions, at a scale that was not previously possible.
While a large portion of these interactions are casual, some interactions carry useful topical information.
There has been a growing interest in identifying \textit{topical experts} on social media using this information \cite{guy2013mining,cheng2013peckalytics,zhao2013topic,bi2014experts}. 
While most studies focus on identifying the top experts in a topic, an under-explored area of research is to rank all users by their topical expertise.
Identifying and ranking intermediate topical experts, in addition to the top experts, can provide great value to users via applications that involve question-answering, crowd-sourced opinions, recommendation systems and influencer marketing.


Expertise may manifest itself via different data sources on different social networks. 
Leveraging information from multiple social networks, and combining them with webpage metadata, can lead to a more complete understanding of users' expertise, compared to using any single source.
Additionally, a scalable solution to such a problem must be able to process hundreds of millions of users and identify experts for thousands of topics.

Our contributions in this study are as follows:

1. \textbf{Feature Diversity:} We present a comprehensive set of \textbf{37 features} that indicate topical expertise on social networks, and provide an in-depth analysis of their predictability and coverage. 
We derive these features from more than \textbf{12 billion message texts}, \textbf{23 million Twitter Lists}, \textbf{58 billion social graph edges}, \textbf{1 million webpages} and \textbf{20,000 Wikipedia pages}.

2. \textbf{User and Topic Coverage:} 
We rank over \textbf{650 million experts} with varying levels of topical expertise across \textbf{9,000 topics}, including popular as well as niche topics.

3. \textbf{Multiple Networks:} We incorporate data from four major social networking platforms: Twitter (TW), Facebook (FB), Google+ (GP) and LinkedIn (LI), and combine it with data from Wikipedia (WIKI) and internet webpage text and metadata.

4. \textbf{Evaluation:} We evaluate the features on a ground truth dataset containing almost \textbf{90,000 labels}.
As far as we know, this is one of the largest datasets used for evaluating topical expertise.

5. \textbf{Open Data:} We make the rankings for top experts and expertise topics for a Twitter user available through \textbf{open public APIs}. 
We also make the topic ontology available as an open dataset. 
We perform our study and analysis on a full production system available on the Klout platform.
Klout\footnote{Klout platform is a part of Lithium Technologies, Inc.} is a social media platform that aggregates and analyzes data from social networks like Twitter, Facebook, Google+, LinkedIn. 

\section{Related Work}
\label{section:related_work}
A growing body of academic as well as industrial research has focused on mining topical experts \cite{popescu2013mining,campbell2003expertise}. 
In \cite{guy2013mining,kolari2008expert,ehrlich2008searching}, enterprise users were studied.
Among social media platforms, Twitter has been undoubtedly the most studied due to the large volume and public nature of its data.
The work in Cognos \cite{ghosh2012cognos} presented a system utilizing only Twitter-List features
and results showed an improvement over Twitter's in-house system \cite{gupta2013wtf}.
In our system, we generate features based on users added to lists as well as on which lists a user creates or subscribes to.
Twitterrank \cite{weng2010twitterrank} presented a system that identifies topical influencers based on the link structure shared between users and on the information extracted from the `bio' in the Twitter profile.
The work in \cite{pal2011identifying} presents a multi-feature approach towards expert mining.
They use a set of 15 features for characterizing social media authors based on both nodal and topical properties and present results across 20 different topics.
The work by LinkedIn \cite{Thuc2014expertise} has focused on determining topic experts based on a LinkedIn skills data set available to the company internally.
Other works have focused solely on niche topics: \cite{bi2014experts} has focused on mining experts in landscape photography and \cite{zhang2007expertise} analyzed Java-forums to identify experts in the Java community.

Multiple works have proposed applications that utilize a topic expert list. 
Cognos and Twitter's Who To Follow service focused on services to recommend experts for the purpose of `following'. 
Research work \cite{wu2011says} has also utilized experts as a starting set to study other social media properties like extent of homophily within categories, speed of information flow and content lifespan.
Works have been presented that utilize the community around topics to understand top influences within a topic \cite{bi2014scalable}. 
Topical expertise of popular users has also been used to mine topics of interest for other users \cite{bhattacharya2014inferring}.
To understand organizational experts, work in \cite{fu2007finding} proposed using a seed of already identified top experts and then followed the network graph to identify other potential experts.
Many works have focused on question-answering services where an expert system provided the core platform to route questions and match experts to askers \cite{zhao2013topic, liu2005finding, jurczyk2007discovering,adamic2008knowledge}.

\section{Problem Setting}
\label{section:problem_setting}
We identify experts in topical domains as those users who produce and share topical information which is recognized as relevant and reliable by other users in the network.
We aim to capture topical interactions that indicate expertise, and derive features for them.
We examine how these features from different sources behave across the global population of users.

To gain insights into the performance of each feature, we use user in-degrees in the graph as the comparison metric.
Since we are dealing with multiple networks, we combine degree information of a user on different graphs into a single quantity that we call \textit{connectivity}.
For a user $u$, we define the \textit{connectivity}, $C_u$, as:
\begin{equation} \label{equation:connectivity}
C_u = ||\mathbf{c_u}||
\end{equation}
where $\mathbf{c_u} = [c_u^1, c_u^2, ..., c_u^n]$ is the connectivity vector and each element $c_u^k$ is the in-degree for the user in the network $k$, e.g. Twitter Followers, Facebook Friends, LinkedIn Connections.
The connectivity may vary from single digit edges for passive users, to hundreds of millions for celebrities, capturing the entire gamut of users.

\begin{figure}[htbp]
  \vspace{-0.1in}
  \begin{center}
    \includegraphics[width=0.8\textwidth,natwidth=610,natheight=642]{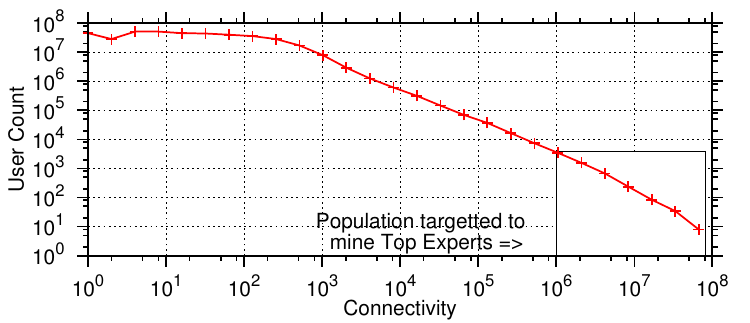}
  \end{center}
  \caption{User Count Distribution vs Connectivity}
  \label{figure:connectivity_distribution}
  \vspace{-0.15in}
\end{figure}

In Fig.~\ref{figure:connectivity_distribution}, we plot the number of users against connectivity.
We observe a distribution where a large number of users have small connectivity and only a small fraction of users have very large values of connectivity.
Most results from previous works like \cite{ghosh2012cognos,pal2011identifying} that find top experts are mainly applicable to the head of this distribution.
Here we instead aim to rank experts over the full distribution, i.e,~on the head as well as the long-tail of expertise.

\subsection{Problem Statement}
\label{section:problem_statement}
At Klout, topics are represented as entries in a hierarchical ontology tree, $\mathcal{T}$.
The tree structure has three levels: \textit{super}, \textit{sub} and \textit{entity}, each with $15$, $602$ and $8,551$ topics respectively.
More details are available in our earlier work \cite{nemanja-lasta}.

We wish to compute an expert score for each user-topic pair.
To begin, we define a feature vector $\mathcal{F}(u, t_i)$ for a user $u$ and a topic $t_i$ as: 
\begin{equation} \label{equation:feature_vector}
\mathcal{F}(u, t_i) = [f_1(u, t_i), f_2(u, t_i), ..., f_m(u, t_i)]
\end{equation} where $f_k(u, t_i)$ is the feature value associated with a specific feature $f_k$.
The normalized feature values are denoted by $\hat{f_{k}}(u, t_i)$ and the normalized feature vector is represented as: $ \hat{\mathcal{F}}(u, t_i) = [\hat{f_{1}}(u, t_i), \hat{f_{2}}(u, t_i), ..., \hat{f_{m}}(u, t_i)].$

The expert score for a user-topic pair denoted by $\mathcal{E}(u, t_i)$ is computed as the result of the dot product of a weight vector $\mathbf{w}$ and the normalized feature vector, $\hat{\mathcal{F}}(u, t_i)$. 
\begin{equation}\label{equation:expert_score}
\mathcal{E}(u, t_i) = \mathbf{w} \cdot \hat{\mathcal{F}}(u, t_i)
\end{equation}
The weight vector is computed with supervised learning techniques, using labeled ground truth data.

Our ground truth data collection system generates user-pair labels across multiple topics.
For a topic $t_i$, the labeled data between two users is defined as:
\begin{equation} \label{equation:gt_label}
 label(u_1, u_2, t_i) = \begin{dcases*}
        +1 & if $u_1$ is voted up, \\
        -1 & if $u_2$ is voted up.
        \end{dcases*}
\end{equation}
For a topic $t_i$ and two users $u_1$ and $u_2$, we define the feature difference vector as:
\begin{equation} \label{equation:feature_delta}
\hat{\mathcal{F}}_{\Delta}(u_1, u_2, t_i) = \hat{\mathcal{F}}(u_1, t_i) - \hat{\mathcal{F}}(u_2, t_i)
\end{equation}
For a feature $f_k$, the corresponding element in the vector $\hat{\mathcal{F}}_{\Delta}(u_1, u_2, t_i)$ is represented as $\hat{f}_{k\Delta}(u_1, u_2, t_i)$.
We can thus compare the expertise of a user-pair by operating on their feature difference vector $\hat{\mathcal{F}}_{\Delta}(u_1, u_2, t_i)$.


\subsection{System Details}
\label{section:system_detail}

\begin{figure}[htbp]
 \centering
 \fbox{\includegraphics[width=0.6\columnwidth,natwidth=610,natheight=642]{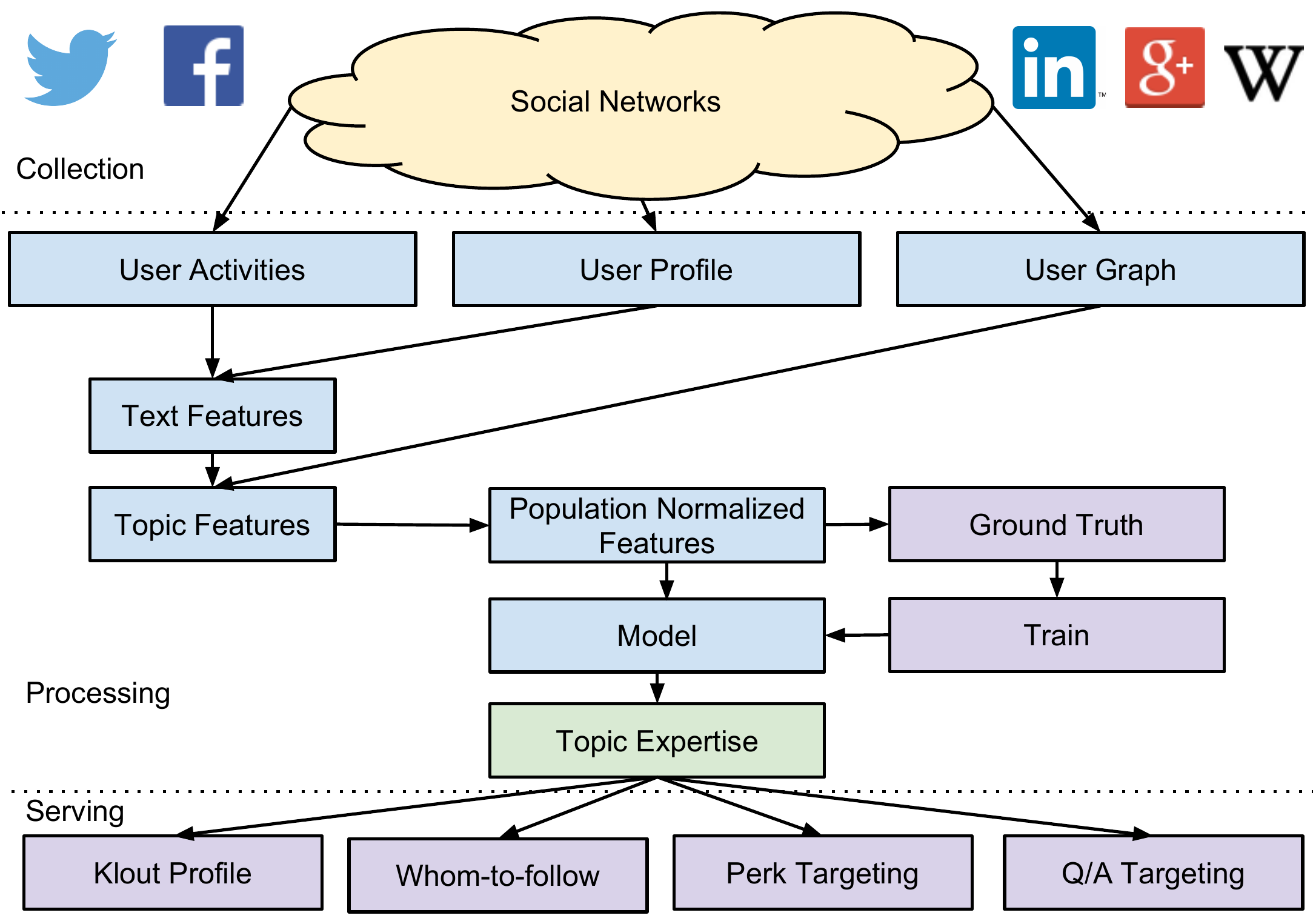}}
 \caption{Topic Expertise Pipeline}
 \label{figure:topics_expertise_overview}
 \vspace{-0.15in}
\end{figure}

Fig.~\ref{figure:topics_expertise_overview} presents an overview of our production system. 
When a user registers on Klout, he connects one or more social networks, and grants permission to Klout to collect and analyze his data.
We use oauth-tokens provided by registered users to collect data from Facebook, LinkedIn and Google+.
We also collect data about users' Twitter graph and list-mentions using oauth-tokens. 
Klout partners with GNIP to collect public data available via the Twitter Mention Stream\footnote{\url{http://support.gnip.com/sources/twitter/overview.html}}.
We do not distinguish between human, organizational and non-human social media accounts and refer to all accounts as users in the rest of the paper. 

We collect, parse, extract and normalize features for hundreds of millions of users daily.
We use scalable infrastructure in the form of Hadoop MapReduce and Hive to bulk process large amounts of data.
The daily resource usage for uncompressed HDFS data reads and writes during feature generation is: 55.42 CPU days, 6.66 TB reads, 2.33 TB writes.

\section{Ground Truth}
\label{section:ground_truth}

\begin{figure}[htbp]
 \centering
 \includegraphics[width=\columnwidth,natwidth=360,natheight=216]{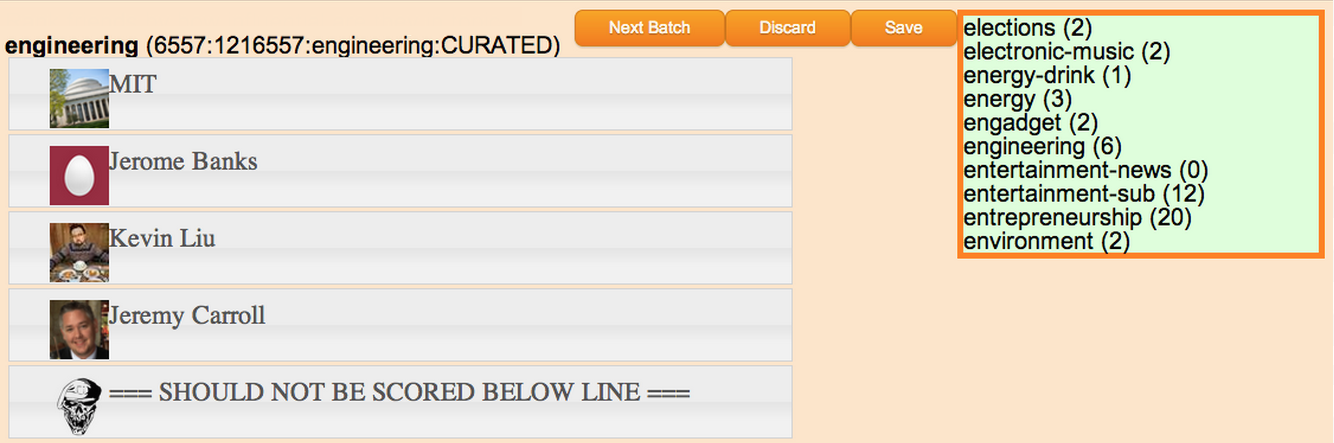}
 \caption{Ground Truth Tool Web UI}
 \label{fig:topics_rank_tool}
 \vspace{-0.09in}
\end{figure}

Ground truth data was collected in a controlled experiment via trusted internal evaluators. 
The data collection tool web UI is shown in Fig. \ref{fig:topics_rank_tool}.

All evaluators were given guidelines on how to use the tool, but no specifics on how to interpret expertise.
For a wide range of topics, evaluators were shown a user list asked to sort the users in order of expertise.
Evaluators also had the option to mark a user as un-sortable or irrelevant to the topic.
To reduce ambiguity in the data, an evaluator only judged users whom he or she was familiar with, and a personalized evaluation dataset was created for each evaluator.
The dataset included selected users from an evaluator's outgoing social graph such as Facebook Friends or Twitter Following; or users with whom the evaluator had interacted through Facebook comments, Twitter retweets and so on.
Users from the dataset were placed in an evaluator's topic list if at least one of their features had a non-zero value for the given topic.
Table \ref{table:gt_statistics} presents details about the collected dataset. 

\begin{table}[htbp]
\caption{Ground Truth Statistics}
\begin{center}
\begin{tabular}{|l|l|}
\hline
Labels for $(u_i,u_j,t_k)$ & 88,995 \\ \hline
Unique Triplets $(u_i,u_j,t_k)$ & 84,964\\ \hline
Unique User Pairs $(u_i,u_j)$ & 66,656 \\ \hline
Unique Users & 4448 \\ \hline
Average List Length & 8.19 \\ \hline
Unique Topics & 746 \\ \hline
Number of Evaluators & 38 \\ \hline
\end{tabular}
\end{center}
\label{table:gt_statistics}
\vspace{-0.2in}
\end{table}


For each topic $t$ and evaluator $e$ with input user list $U_{in}(e, t)$, we obtain the sorted list
$U_{sorted}(e, t)$, which excludes the set of un-sortable users.
The sorted list is represented as: $U_{sorted}(e, t) = [u^s_1, u^s_2, ... , u^s_N]$. 
Thus, as per the evaluator, $\mathcal{E}(u^s_i, t) > \mathcal{E}(u^s_j, t)$, for $i,j \in {(1..N)}, i > j$.

Unique user-pairs $(u_i, u_j)$ are created from the quadratic explosion of $U_{sorted}$, with $i,j \in {(1..N)}, i > j$.
%
%
A label (+1, -1) is generated for the pair $(u_i, u_j)$ as described in Eq.\ref{equation:gt_label}, depending on the users' relative position on the list.
The training and data evaluation is performed on $\hat{\mathcal{F}}_{\Delta}(u_i, u_j)$ for each $(u_i, u_j)$.



\label{subsec:ground_truth_collection}



\begin{figure}[htbp]
  \centering
  \includegraphics[trim=0 10 0 25, clip, width=0.65\columnwidth,natwidth=360,natheight=216]{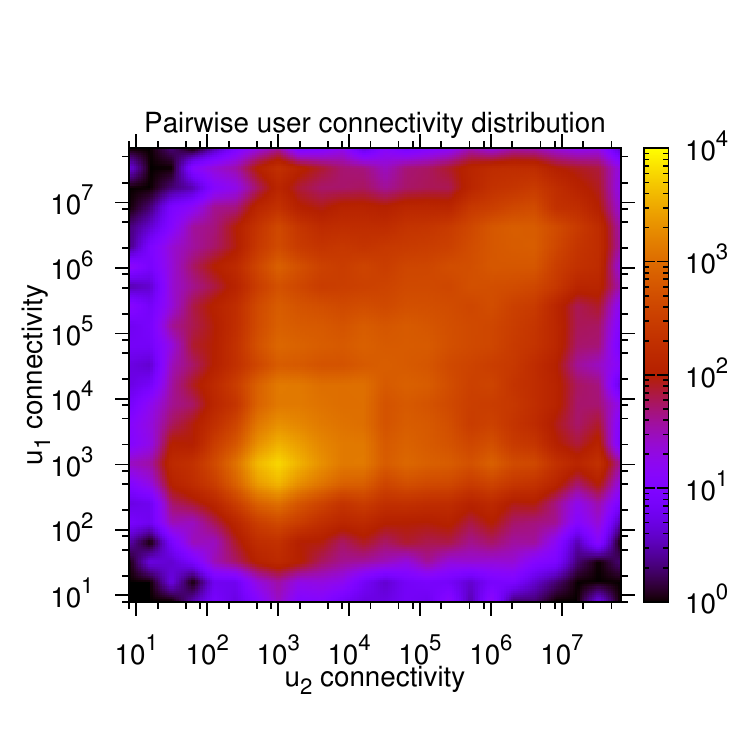}
  \caption{Ground Truth Connectivity Distribution}
  \label{figure:gt_connectivity_distribution}
  \vspace{-0.1in}
\end{figure}

Since our goal is to study the feature distribution for the entire user population, it is important we collect ground truth for users with a wide range of connectivity.
In Fig. \ref{figure:gt_connectivity_distribution}, we plot the connectivity distribution for users in the ground truth dataset.
The heatmap represents the numbers of users for a given pair of connectivity.
For users with connectivity between $10^2$ and $10^7$, we observe that we have a good coverage in the number of evaluations, with the most evaluations for users between $10^2$ and $10^4$.

\begin{figure}
\centering
\begin{subfigure}{.56\textwidth}
  \centering
  \includegraphics[width=0.95\textwidth,natwidth=360,natheight=216]{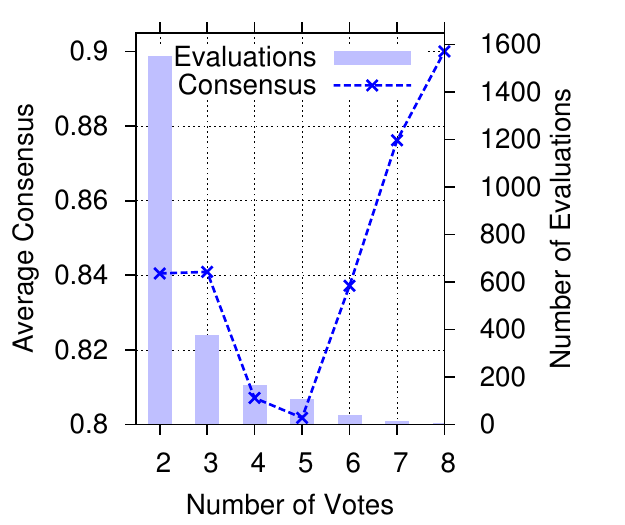}
  \caption{Number of Votes Casted}
  \label{fig:gt_agreement_distribution}
\end{subfigure}%
\begin{subfigure}{.45\textwidth}
  \centering
  \includegraphics[width=0.95\textwidth,natwidth=360,natheight=216]{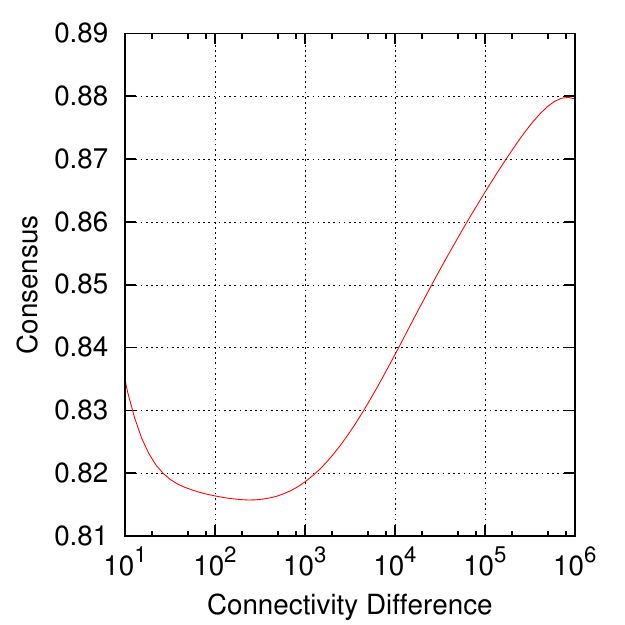}
  \caption{Connectivity Difference}
  \label{fig:gt_connectivity_delta_agreement_distribution}
\end{subfigure}
\caption{Label Consensus Distribution}
\label{fig:consensus_distribution}
\vspace{-0.12in}
\end{figure}

Since expertise is somewhat subjective based on evaluators' perception, a strict definition of expertise was not provided to evaluators. 
We show the label consensus among the human evaluators within the ground truth in Fig. \ref{fig:gt_agreement_distribution}. 
We define label consensus as the number of voted up labels for a pair $P(u^s_i, u^s_j)$ divided by total votes casted for the pair and topic.
One can observe that when two unique evaluators are asked to order a user-pair, in $84\%$ of cases, evaluators agree on the ordering. 
As the number of casted votes grows, consensus drops to the minimum of $80\%$ for $5$ unique evaluators, after which consensus grows to about $90\%$ for $8$ evaluators. 
We can conclude that humans do not fully agree on exact expertise ordering, thereby imposing fundamental limitations on supervised machine learned models.
 
Finally, we present the dependency of connectivity difference between evaluated user-pair on consensus in Fig. \ref{fig:gt_connectivity_delta_agreement_distribution}.
Consensus exhibits an increasing trend from $82\%$ to $88\%$ as connectivity difference increases. 
This is expected as users with higher connectivities might be more easily recognized as experts. 


\section{Feature Analysis}
\label{section:analysis_and_evaluation}


We present 37 features here that capture topical expertise for users from various sources.
The textual inputs from each source are mapped to bags-of-phrases by matching against a dictionary of approximately 2 million phrases.
Bags-of-phrases are then mapped to a topic ontology to create bags-of-topics, reducing the dimensionality of the text from 2 million to more than $9,000$ topics. 
These bags-of-topics are exploded, and for each user-topic pair $(u, t_i)$, we build the feature vector $\mathcal{F}(u, t_i)$.
A more detailed discussion about feature generation is provided in our earlier work \cite{nemanja-lasta}.

Each element in the feature vector, $f_k(u,t_i)$, has a naming convention:
\newline
\texttt{\detokenize{<Network>_<Source>_<Attribution>}}, 
that encodes the following characteristics: (a) the social network of origin, (b) the source data type, and (c) the attribution relation of a feature to the user. 
The networks we consider include Twitter (TW), Facebook (FB), Google+ (GP), LinkedIn (LI) and Wikipedia (WIKI). 
Table \ref{table:feature_names} summarizes the attribution relations and their descriptions.


\begin{table}[htbp]
\caption{Feature Attribution Nomenclature}
\begin{tabular}{p{0.25\columnwidth} p{0.65\columnwidth}}
  \hline
  \multicolumn{2}{c}{\textbf{Attribution}} \\
  \hline
  GENERATED & Originally generated or authored content by the user, including posts, tweets, and profiles. \\ 
  REACTED & Content generated by another user (actor), but as a reaction to content originally authored by the user under consideration. This includes comments, re-tweets, and replies. \\
  CREDITED & Text that is associated with the user without direct involvement from the user. Examples include tags and lists. \\
  GRAPH & Topics aggregated and derived from a user's friends, followers and following users. \\
  \hline
\end{tabular}
\label{table:feature_names}
\vspace{-0.1in}
\end{table}

Below we describe the different source data types considered:
\paragraph{\textbf{Message Text}}
We derive features as the frequency of occurrence of topics in text of messages posted by users, including original posts, comments and replies. 
These data sources are named under `MSG TEXT', and provide useful topical information for all users who are actively posting and reacting to messages on the network.
In the case of FB Fan Pages, we call this data source `PAGE TEXT', to differentiate from message text in personal FB pages.
The feature vectors are thus named as \texttt{\detokenize{TW_MSG_TEXT_GENERATED}}, \texttt{\detokenize{TW_MSG_TEXT_REACTED}}, \texttt{\detokenize{TW_MSG_TEXT_CREDITED}} and so on.

We present the number of message texts processed for each network in Table \ref{table:message_text_processed}.
Since we have to access to the entirety of Twitter's public data, the volume of messages processed from Twitter is significantly more than the number of messages from Facebook.

 

\begin{table}[htbp]
\small 
\caption{Message Text Processed}
\begin{center}
\vspace{-0.05in}
\begin{tabular}{|c|c|}
  \hline
  Network & Number of Message Texts  \\ \hline
  Twitter & 11 Billion \\ \hline
  Facebook & 696 Million \\ \hline
  Facebook Pages & 1.47 Billion \\ \hline
  Google+ & 22 Million \\
  \hline
\end{tabular}
\label{table:message_text_processed}
\vspace{-0.3in}
\end{center}
\end{table}

\paragraph{\textbf{User Lists}}
One of the most important data sources related to topical expertise is user lists on Twitter, as previously explored in \cite{ghosh2012cognos,cheng2013peckalytics}. 
A user may be added to topical lists by other users, thereby marking him or her as an expert. 
We refer to this data source simply as `LIST'.
In addition we also derive other features from the lists a user creates or subscribes to.
The list-based features are derived from over 23 million lists corresponding to over 7.5 million unique users.

Since only a subset of lists per user is available for collection from Twitter, we estimate this feature as $f_{k}(u, t_i) \approx L_{c}(u, t_i) \times \frac{L(u)}{L_{c}(u)}$ where $L_{c}(u, t_i)$ is the number of collected lists for the user on topic $t_i$, $L_{c}(u)$ is the total number of collected lists, and $L(u)$ is the true number of lists for the user, retrieved from the user's Twitter profile.

\begin{figure*}[htbp]
 \centering
 \includegraphics[width=\linewidth,natwidth=360,natheight=216]{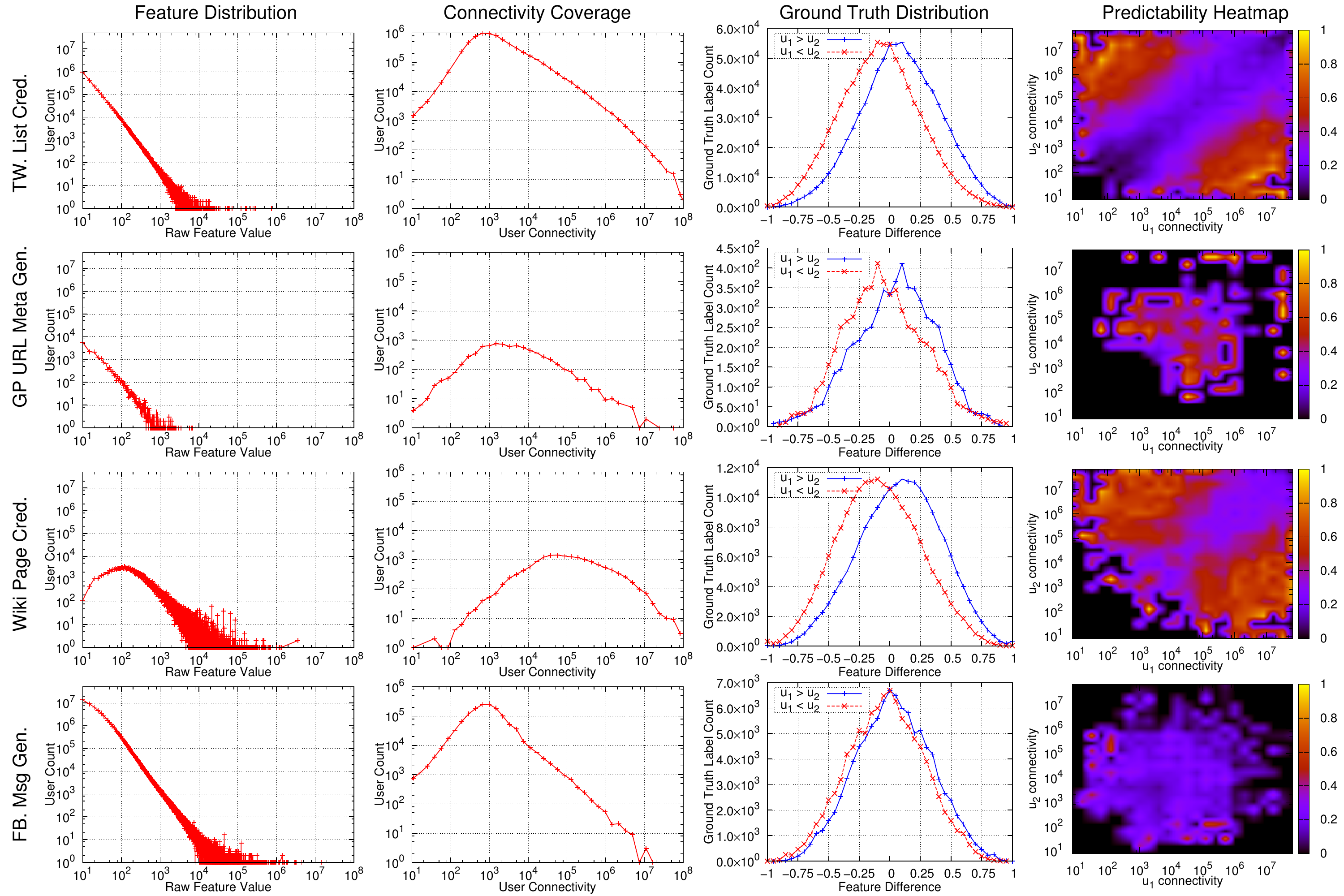}
 \caption{Feature Distribution (FD), Connectivity Coverage (CC), Ground Truth Distribution (GTD), Predictability Heatmap (PH) for selected features}
 \label{fig:distribution_and_predictability}
 \vspace{-0.1in}
\end{figure*}

\paragraph{\textbf{User Profile}}
We derive features based on the user-listed information available in user profiles.
For a given user, we extract the skill and industry information from LinkedIn profiles to derive topic signals.
We assign the number of followers of a company normalized over the number of followers for the particular industry as the feature value.
Feature vectors are thus named as \texttt{\detokenize{LI_SKILLS_GENERATED}}, \texttt{\detokenize{LI_INDUSTRY_GENERATED}}.

\paragraph{\textbf{Social Graphs}}
We leverage graph and peer-based information to derive topical signals from a user's connections. 
For a given user and topic, we aggregate the topic strength across all of his connections, and scale this with the aggregated strength for the global population.
Apart from `FOLLOWERS', these features may come from other graph sets such as `FOLLOWING' and `FRIENDS', depending on the networks.
This feature is especially important for the set of users who may not explicitly talk about certain topics through messages, but may yet be recognized as experts in their fields.
For example, if a large number of experts in \textit{finance} follow Warren Buffett on Twitter, then he can be identified as an expert in the topic, though he may not post messages related to \textit{finance}.
The features are derived from more than 56 billion follower and 2.7 billion following edges from Twitter and 1.69 billion friend edges from Facebook.



\paragraph{\textbf{Webpage Text \& Metadata}}
Users often share and react to URLs on social media that are related to their topics of expertise.
We extract and create features from the content body of such shared URLs, and also from metadata such as categories associated with the shared URLs. 
These data sources, named `URL' and `URL META' respectively, provide rich contextual information about the associated social media message. 

Inversely, articles that are published on online publishing platforms often contain useful metadata, including social identity information for the authors through the Open Graph Protocol and Twitter Creator cards\footnote{\url{http://ogp.me}, \url{https://dev.twitter.com/cards/markup}}.
Such URLs and published articles provide a means of bridging social network information with internet webpage data, thereby enabling a richer understanding of content as it relates to users.
We process 1 million such webpages that were the most shared or reacted over a $90$-day rolling window; extracting features for over 50,000 users based on the popularity of the page.
In this case, we derive the feature value as: 
\[ f_k(u, t_i) = \sum\limits_{j=1}^{N} tf(t_i, BT_{j}^{SWWW}) \times n_j \]
where $N$ is the total number of documents attributed to the user, $tf(t_i, BT_{j}^{SWWW})$ is the topic frequency in the bag of topics derived for the $j$-th document, and $n_j$ is the number of times the $j$-th document was reacted upon in the network.
The features from this data source, which we call `SOCIAL WWW', is especially important to attribute expertise to journalists and bloggers who write long form articles, and can be recognized as topical experts due to their authorship.

In addition to social networks and webpage text and metadata, we also consider Wikipedia data\footnote{https://dumps.wikimedia.org/} that provides information about a user's expertise.
We manually identify mapping for over $20,000$ users from Wikipedia page to the user's social network identity.
The existence of a Wikipedia page for a user itself may be a strong signal that he is recognized for his expertise in some topic.
However, in order to eliminate spurious pages, we do not consider merely the existence of a page to be a signal. 
Instead, we compute the inlink-to-outlink ratio for a Wikipedia page, which indicates the authority of the user's page. 
Using pagerank gives similar results, so we only present results for the inlink-to-outlink feature.
The final feature values are computed as:
\[ f_k(u, t_i) = tf(t_i, BT^{WIKI}(u)) \times \frac{L_{in}}{L_{out}} \]
where $tf(t_i, BT^{WIKI}(u))$ is the topic frequency in the bag-of-topics derived from the Wikipedia page, and $L_{in}$, $L_{out}$ are the number of inlinks and outlinks to the page respectively, in the full Wikipedia graph.


Note that the features described above are applicable for mining topical interests as well, but here we are focused on topical expertise. 
The major difference between the two problems is in feature normalization, where topical expertise features are scaled against the global population, and topical interest features are scaled with respect to the user.
The problem of expertise has to scale to millions of users per topic, while that of interests has to scale to hundreds of topics per user.

\subsection{Feature Distribution}

In this section, we describe four ways we analyze and visualize the coverage and predictability of each feature. 

\subsubsection{\textbf{Feature Distribution}}
The first column in Fig. \ref{fig:distribution_and_predictability} shows the population feature distribution for selected features.
The distributions are plotted on the log-log scale, where the number of users is plotted on the y-axis against the raw feature values on the x-axis. 
Features such as Wikipedia and Google+ URLs are present for less than $10^5$ users, while the other two features are present for a much greater number of users.
We observe that when plotted on the log-log scale, the number of users has an almost linear relationship to the feature values. 
The plots suggest that most features could be modeled as power law distributions over the population of users under consideration.
We therefore rescale the features by transforming them as follows:
\begin{equation}
\hat{f_{k}}(u, t_i) = \frac {log(f_{k}(u, t_i))}{\operatorname*{max}\limits_{u_i \in U} log(f_{k}(u_i, t_i))}
\end{equation}

\subsubsection{\textbf{Feature Connectivity Distribution}}
The second column in Fig. \ref{fig:distribution_and_predictability} plots the number of users who possess the specified feature against the connectivity of the users. 
The Twitter List feature is present over the entire range of users, but Wikipedia is present only for users with high connectivity. 
The URL feature for Google+ on the other hand is present for only a small number of users with low to medium connectivity.
These differences in coverage highlight the need to have features that can capture information for different sections of the population.

\begin{table*}[htbp]
\caption{\textbf{Feature Catalog}: Precision (P), Recall(R), F1 Score, Coverage [\% of corpus] (C), Feature Distribution (FD), Connectivity Coverage (CC), Ground Truth Distribution (GTD), Predictability Heatmap (.).
The axes for Feature Distribution, Connectivity Coverage, Ground Truth Distribution and Predictability Heatmap are equivalent to the axes in Fig. \ref{fig:distribution_and_predictability}.
}
\begin{center}
\includegraphics[width=\textwidth,natwidth=622,natheight=1029]{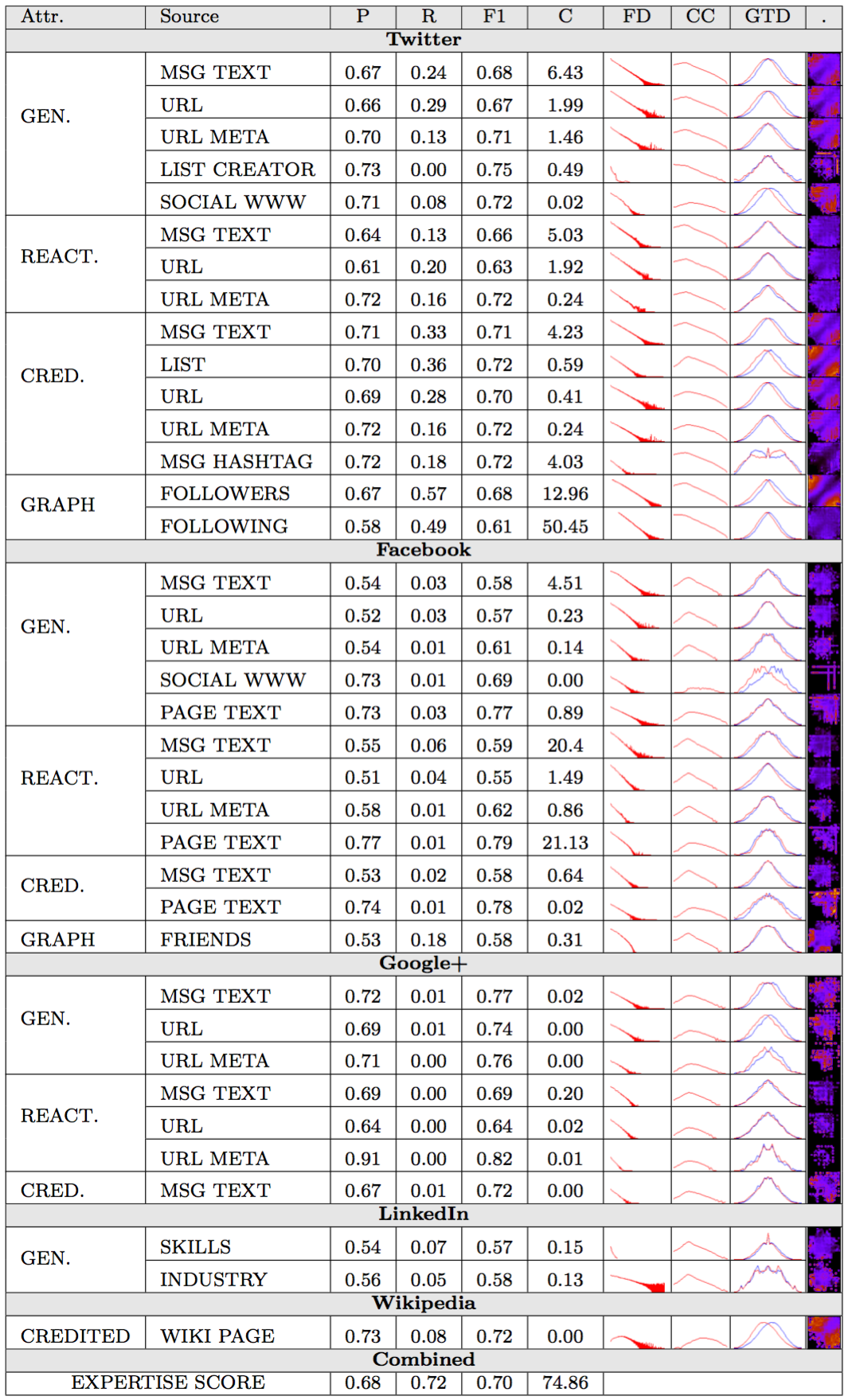}
\end{center}
\label{table:features}
\vspace{-0.2in}
\end{table*}%

\subsubsection{\textbf{Ground Truth Distribution}}
Fig. \ref{fig:distribution_and_predictability} also shows the ground truth distribution for the features in the third column. 
In this case, the number of labels in the ground truth are plotted against the normalized feature difference values between user pairs. 
The labels plotted had at least one user in the pair possessing a non-zero feature value. 
For a user-pair $(u_1, u_2)$ we plot $\hat{f}_{k\Delta}(u_1, u_2, t_i)$ and $\hat{f}_{k\Delta}(u_2, u_1, t_i)$, allowing us to symmetrically visualize the differentiating nature of the feature over the ground truth. 
We observe that for Wikipedia, Twitter Lists and Google+ URLs, the curves show greater separation compared to Facebook message text, which has almost overlapping curves. 
This shows that the former features have a higher ability to identify experts compared to the features derived from Facebook message text.

\subsubsection{\textbf{Predictability Heatmap}}
In the final column of Fig. \ref{fig:distribution_and_predictability} we observe feature predictability with respect to user connectivity, visualized as a heatmap.
For the labeled pairs in the ground truth, we plot the average feature value difference for the instances when the feature correctly predicted the greater expert in the pair.
The co-ordinates of each point are the connectivities of the first and the second user in a pair, and the brightness of the point indicates the average absolute value of the feature difference.

We observe that the Wikipedia and Twitter List features show a high ability to predict when the connectivity difference between the users is high.
For Twitter Lists, connectivity coverage in the second plot of the same row shows that a majority of users have connectivity of less than $10^5$, but the heatmap shows low predictability for this cohort of users.
Twitter Lists thus have weaker predictability in the long tail of the global population distribution, but still provide significant coverage to scale our systems and are good predictors for top experts.

For moderately connected users and users with similar connectivities, Google+ URLs feature is a good predictor, and behaves in a complementary manner to Twitter Lists and Wikipedia.
Facebook messages show poor predictability, which may be because Facebook messages are more conversational and not very indicative of topical expertise.
Note that other Facebook features do have better predictability, but here we highlight this feature for contrast and comparison.



\subsection{Feature Catalog}

Table \ref{table:features} shows the precision, recall, F1 scores and the population coverage percentage for the full list of features used. 
The features are evaluated over the ground truth data, where the prediction by a feature $f_k$ is correct when the following relation holds:
\begin{equation}
label(u_1, u_2, t_i) \cdot sgn(\hat{f}_{k\Delta}(u_1, u_2, t_i)) > 0,
\end{equation}
where $sgn(x)$ is the signum function.


We observe from Table \ref{table:features} that the Twitter List based features have some of the highest precision values on the ground truth data. 
This corroborates the approach used in \cite{ghosh2012cognos}, where Twitter Lists are used to identify experts. 
The SOCIAL WWW and Facebook Pages features show similar behavior to Twitter Lists, with high precision and F1 scores, but low coverage and recall values.
Other features such as those derived from `Credited' message texts and hashtags, and those derived from the followers of a user, prove to have high precision, F1 score, and population coverage.



Features based on URLs, URL META and message text from Google+ have higher precision and F1 scores than Facebook and some Twitter features. 
One reason for this could be that users tend to post messages of greater length on Google+ compared to other networks \cite{nemanja-lasta}. 
For LinkedIn, though the precision and recall values are low, we observe from the heatmaps that they have high predictability for users with low connectivity, indicated by the bright spots near the lower left corners.
Therefore, they are still useful in identifying experts in the long tail.
Finally, Since well-known personalities with Wikipedia pages are already recognized as experts in their domains, Wikipedia features have high precision and F1 scores.

To conclude, the best features for topical expertise in terms of F1-scores are Twitter Lists, Wikipedia pages, Social WWW, Facebook Fan Page text and URL metadata features. Graph and text based features provide long tail coverage, though they are less predictive. 

\subsection{Connectivity Analysis}

\begin{figure}[htbp]
 \centering
 \includegraphics[width=0.6\columnwidth,natwidth=360,natheight=216]{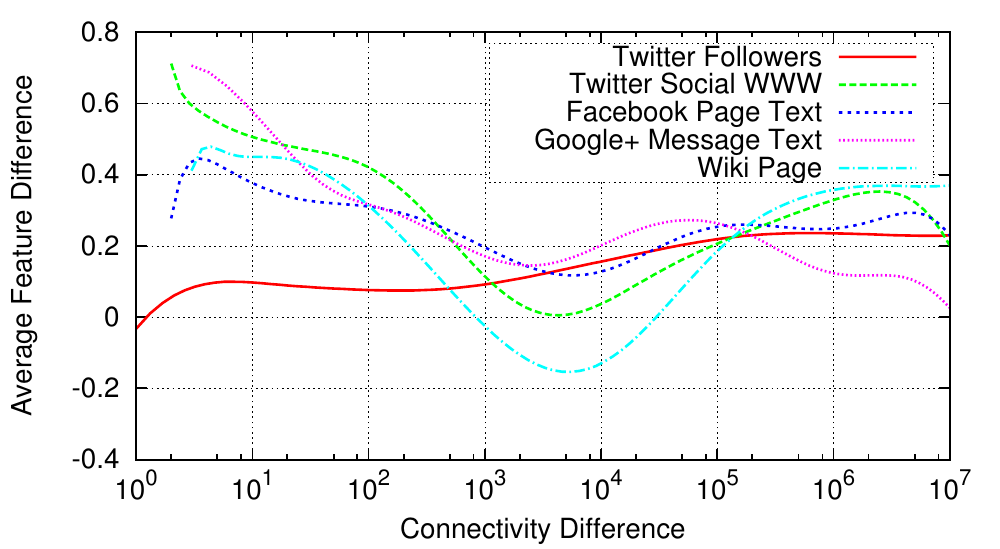}
 \caption{Feature vs Connectivity}
 \label{fig:feature_predictability_by_connectivity}
 \vspace{-0.2in}
\end{figure}

As described in Section \ref{section:ground_truth}, our ground truth dataset contains pairwise comparisons for users with varying amounts of connectivity. 
To study feature behavior across the spectrum of users, we plot the average feature difference 
in a user pair against the connectivity difference, for a few selected features in Fig. \ref{fig:feature_predictability_by_connectivity}.
This is effectively a lower dimensional version of the predictability heatmaps that reveal new insights.

We observe that a few features shown have high predictability when the connectivity difference is low. 
This indicates that even when user network sizes are similar, their expertise can be differentiated with these features.
The predictability for Twitter follower feature increases monotonically with connectivity difference.
Finally, the SOCIAL WWW and Wikipedia features suffer from significant dips in the mid-range, and perform the best when comparing users with network size difference of $10^5$ and above.

\section{Expertise Score}
\label{section:score}
The features described above are combined into a feature vector for each user.
The ground truth dataset is split into training and test sets. 
The labels for the pairs in the training set are fit against the feature difference vectors $\hat{\mathcal{F}}_{\Delta}(u_1, u_2, t_i)$ using Non Negative Least Squares (NNLS) regression.
We constrain the weights to be non negative because the features contributing to the score are designed to be directly proportional to expertise.

Model building is performed in a two step process. 
In the first step we build network level models, based on the disjoint sets of features for each given network.
In the second step we build a global model treating the network expertise scores obtained in the first step as features.
Weighted network models are generated by multiplying the network level weights with their corresponding weight from the global model.
These weighted network models are then combined by concatenation into a single weight vector $\mathbf{w}$.
This two-step process enables representation of sparse features derived from low population networks.
The final expertise score for a user is computed by applying the weight vector to the user's normalized feature vector $\hat{\mathcal{F}}(u, t_i)$ as shown in Equation \ref{equation:expert_score}.

The F measure for the expertise score is $0.70$, covering $75\%$ of users across all networks. 
Table \ref{table:features} shows additional performance metrics.
Given that the human evaluator consensus is close to $84\%$, this F measure is reasonably high.

\subsection{Super-topics distribution}

\begin{table*}[ht!]
\caption{Super-topic user distribution across different networks}
\label{table:super_topic_distribution}
\includegraphics[width=\textwidth,natwidth=360,natheight=216]{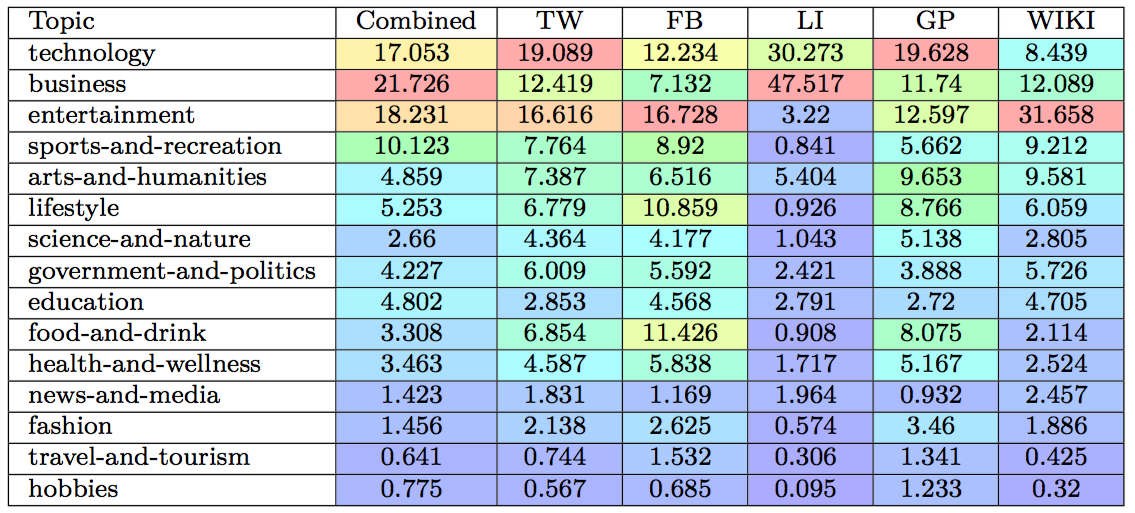}
\end{table*}

In Table \ref{table:super_topic_distribution}, we examine the characteristics of topic expertise aggregated across users on different networks. 
We roll up entities and sub-topics to super-topics to aid the visualization and reduce the topic dimension space from $9, 000$ topical domains to $15$.
We plot percentage breakdown of super-topics on each network for the users on that network and also across all networks combined.
We observe that users in each network have distinct topical expertise distribution.
On Twitter and Google+ the super-topic `technology' is the most represented one, whereas `entertainment' is the most represented super-topic on Facebook. 
Facebook users are also experts in `technology', `lifestyle' and `food-and-drink'.
On LinkedIn, most users are expert in `business' and `technology' and the representation of other super-topics drop-off significantly.
This is expected as LinkedIn is a professional networking platform.
The left-most column shows the distribution of expertise when all the network features are combined to build the combined expertise for users.
The most represented topics are `business', `entertainment' and `technology'.
We also observe that due to our multi-network approach, topical distribution across the users is not as skewed as the distribution for individual networks.

\section{Applications}
\label{section:application}
\subsection{Application Validation}

\begin{figure}
    \begin{subfigure}{0.46\textwidth}
        \centering
        \includegraphics[width=33mm,natwidth=360,natheight=216]{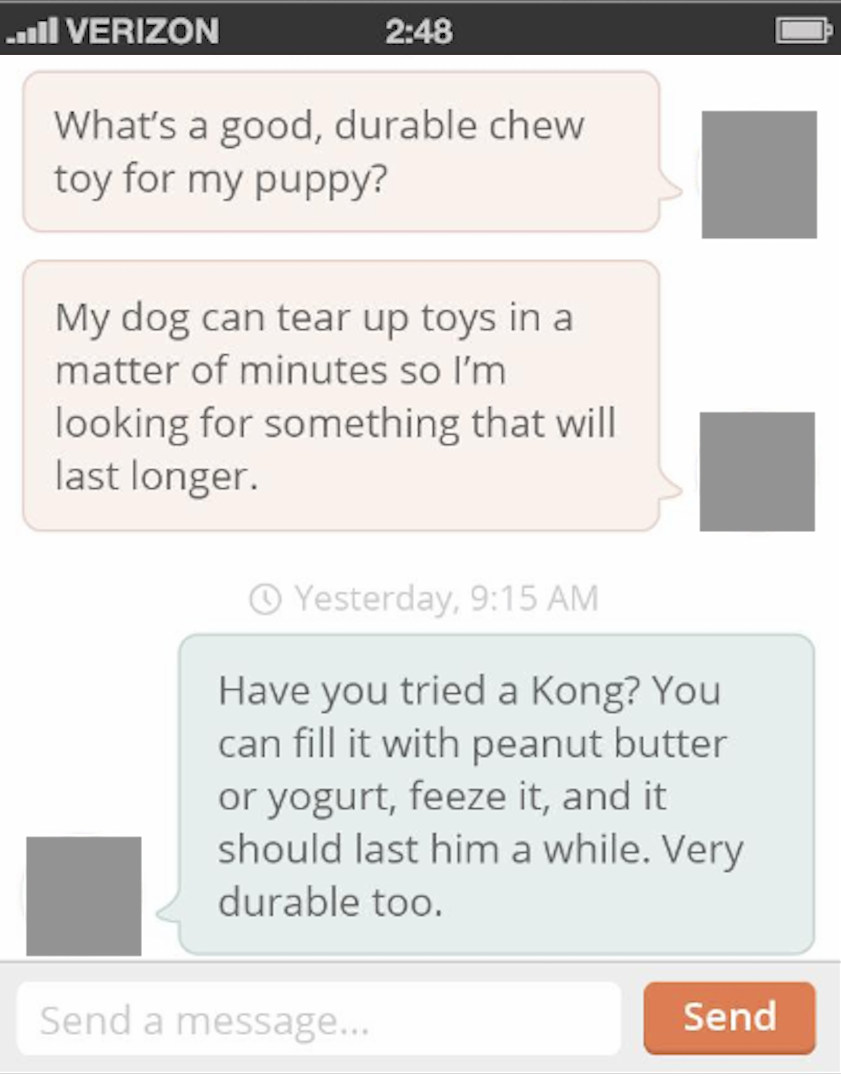}
        \caption{Q \& A application}
        \label{fig:cinch_screenshot}        
    \end{subfigure}%
    \begin{subfigure}{0.50\textwidth}
        \centering
        \includegraphics[width=35mm,natwidth=360,natheight=216]{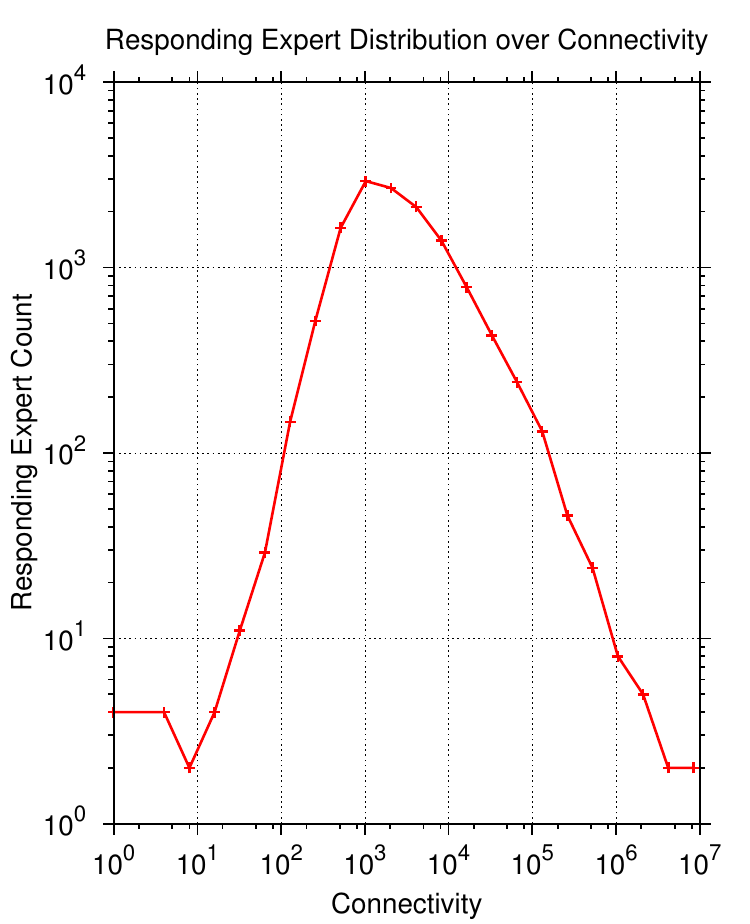}
        \caption{Distribution of responding experts}
        \label{fig:advisor_connectivity}
    \end{subfigure}
    \caption{Q \& A experts}
    \label{fig:q_and_a_experts}
\end{figure}

To validate the expertise system, we developed an application\footnote{Cinch App from Klout, \url{http://tnw.to/t3FtH}} where users could ask questions pertaining to their topics of interest. 
The questions were then routed to topical experts who could presumably answer the questions. 
The topical experts then could choose to respond to the questions posed to them, through a direct conversation with the user asking the question.
An example of such an interaction in the application is shown in Fig. \ref{fig:cinch_screenshot}.

The application generated over $30,000$ such conversations, with more than $13,000$ topical experts responding to questions asked.
Fig. \ref{fig:advisor_connectivity} shows the connectivity distribution for the responding experts.
We observe that a majority of experts who responded fell within the connectivity range of $500$ to $5,000$, whereas top experts with connectivity values of more than $1,000,000$ were rarely responsive.
This shows the value of ranking users who are not necessarily the top experts in their fields while building applications used by the average population.

\subsection{Expert Examples}
In Table \ref{table:feature_examples}, we present user examples with feature values for selected Twitter features and topics, along with their user population counts.
For a topic such as `technology', in a typical $top$-$k$ approach, users like \textit{Google} and \textit{TechReview} (MIT Technology Review) may be regarded as the top experts.
But by ranking the full population, users like \textit{Tim O'Reilly} and \textit{Arrington} with lower feature values can also be identified as experts in the topic.

The effectiveness of a multi-source approach is evident in the examples of `politics' and `machine learning'.
For `politics', \textit{BarackObama} has the highest feature values for the Twitter follower graph feature, due to the large number of users who follow President Obama on Twitter.
However, \textit{Politico} and \textit{NYTimes} have high feature values for other features such as URL and SOCIAL WWW, enabling their identification as experts.
As a niche topic, `machine learning' attracts a relatively small community of users on the social platforms.
For the most socially engaging accounts like \textit{KDnuggets} and \textit{Kaggle}, we observe that many non-zero features exist and it is relatively simple to identify these accounts as top experts.
However, for passive users such as \textit{ylecun}, the small individual feature values add up to recognize him as an expert.

\begin{table*}[htbp]
\begin{center}
\caption{Examples for Twitter Normalized Features}
\label{table:feature_examples}
\includegraphics[width=\textwidth,natwidth=730,natheight=813]{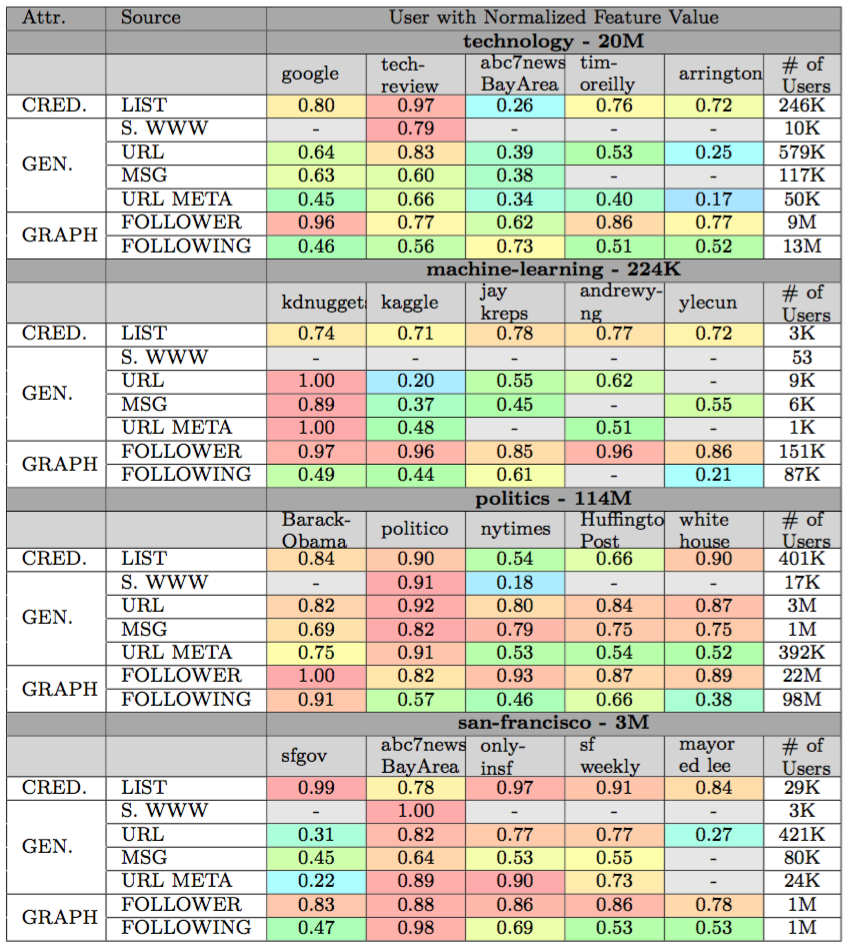}
\end{center}
\end{table*}

\begin{figure*}[htp]
  \centering
  \fbox{\includegraphics[width=\textwidth,natwidth=360,natheight=216]{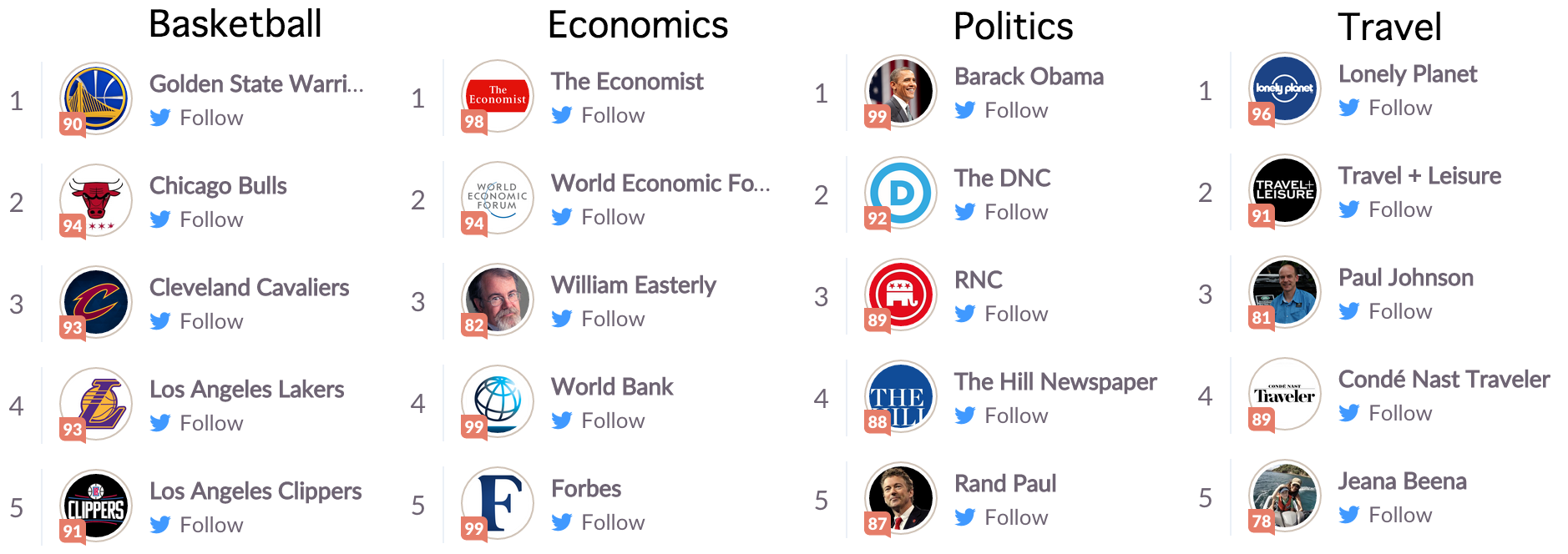}}
  \caption{Top Ranked Experts by Topic; Snapshot on July 12, 2015.}
  \label{fig:topic_influencer}
\end{figure*}

In terms of coverage, the system scores $224$K users for a niche topic like `machine learning',  and approximately~3 million users for `san francisco', a number that is close to half of the bay area's population.
Finally, Fig.~\ref{fig:topic_influencer} shows a screen-shot of the top ranked users for a few selected topics.

\section{Open API}
\label{section:open_api}
We have opened the results of this paper via two REST API endpoints.
The first endpoint returns the top experts for a specified topic.
The second returns the expertise topics for a specified user.
The documentation for using the REST APIs to retrieve these results is available at \textbf{\url{http://klout.com/s/developers/research}}.
The results are provided in JSON format.
The ontology of the $9,000$ topics used is available at \url{https://github.com/klout/opendata}.

\newpage

An example of the API response for top experts in `politics' is provided in Listing \ref{listing:ranked_list_by_topic}.


\begin{lstlisting}[language=json,firstnumber=1,
caption=Ranked experts for `politics',
label=listing:ranked_list_by_topic]
{ 
 "topicSlug": "politics",
 "experts": [
  {
   "twitterUsername": "BarackObama",
   "rank": 1
  },
  {
   "twitterUsername": "washingtonpost", 
   "rank": 2
  }
}
\end{lstlisting}

Listing \ref{listing:topic_score_by_user} presents an example response for the expertise topics for `@washingtonpost'.
The score associated with a topic in the result is the percentile based on the expertise scores within the topic.

\begin{lstlisting}[language=json,firstnumber=1,
caption=Topical expertise percentiles for `washingtonpost',
label=listing:topic_score_by_user]
{ 
 "twitterUsername": "washingtonpost",
 "topicSetType": "expertise",
 "topicSet": [
  {
   "topicId": "7192018630863038134",
   "topicSlug": "journalism",
   "topicDisplayName": "Journalism",
   "topicScore": 0.9999999874118743
  },
  {
   "topicId": "8119426466902417284",
   "topicSlug": "politics",
   "topicDisplayName": "Politics",
   "topicScore": 0.9999999230887799
   }
  ]
}
\end{lstlisting}

\section{Conclusion and Future Work}
\label{section:conclusion}
In this study, we derive and examine a variety of features that predict online topical expertise for the full spectrum of users on multiple social networks.
We evaluate these features on a large ground truth dataset containing almost $90,000$ labels. 
We train models and derive an expertise scores for over 650 million users, and make the lists of top experts available via APIs for more than 9,000 topics.

We find that features that are derived from Twitter Lists, Facebook Fan Pages, Wikipedia and webpage text and metadata are able to predict expertise very well. 
Other features with higher coverage such as those derived from Facebook Message Text and the Twitter Follower graph enables us to find experts in the long tail.
We also found that combining social network information with Wikipedia and webpage data can prove to be very valuable for expert mining. 
Thus a combination of multiple features that complement each other in terms of predictability and coverage yields the best results.

Further studies in this direction could unify cross-platform online expertise information, in addition to other data sources such as Freebase and IMDB.
Another area that could be explored in the future is the overlap and differences in the dual problems of topical expertise and topical interest mining.
To conclude, we provide an in-depth comparison of topical data sources and features in this study, which we hope will prove valuable to the community when building comprehensive expert systems.

\section{Acknowledgements}
\label{section:acknowledgements}

We thank Sarah Ellinger and Tyler Singletary for their valuable contributions towards this study.

%
\bibliographystyle{unsrt}
\bibliography{bibliography} 
%

\end{document}